\def\nmg{{Ni$_2$MnGa}}
\def\e{{$E_F$}}

\documentclass[twocolumn,secnumarabic, graphics,floatfix,nofootinbib,nobibnotes,aps,prb,10pt]{revtex4-1}
\usepackage[english]{babel}
\usepackage{epsf}
\usepackage{epsfig}
\usepackage{graphicx}
\usepackage{xcolor}
\usepackage{ulem}

\begin{document}
	\title{Bulk Electronic Structure of \nmg\, studied by Density Functional Theory and Hard X-ray Photoelectron Spectroscopy}
	
	\author{Joydipto Bhattacharya$^{1,2,\dagger}$, Pampa Sadhukhan$^{3,\dagger}$, Shuvam Sarkar$^3$, Vipin Kumar Singh$^3$, Andrei Gloskovskii$^4$, Sudipta Roy Barman$^{1}$, and Aparna Chakrabarti$^{1,2}$ }
	\affiliation{$^1$Homi Bhabha National Institute, Training School Complex, Anushakti Nagar, Mumbai 400094, India}
	\affiliation{$^2$Raja Ramanna Centre for Advanced Technology, Indore 452013, India}
	\affiliation{$^3$UGC-DAE Consortium for Scientific Research, Khandwa Road, Indore 452001, India}
	\affiliation{$^4$Deutsches Elektronen-Synchrotron DESY, Notkestrasse 85, D-22607 Hamburg, Germany}
	\affiliation {$^{\dagger}$Both the authors contributed equally.
	}
	
	\begin{abstract}
		A combined study employing density functional theory (DFT) using the experimentally determined  modulated structures in the martensite phase and bulk-sensitive hard x-ray photoelectron spectroscopy (HAXPES) of  stoichiometric single-crystalline \nmg\, is presented in this work. The experimental valence band (VB) features closely match the theoretical VB calculated by DFT using generalized gradient approximation  for both  the martensite and austenite phases. 
		~We establish the existence of a charge density wave (CDW) state in the martensite phase from the shape of the VB near the Fermi level (\e). This shows (i) a transfer of spectral weight from the near \e\, region to the higher binding energy side resulting in a dip-peak structure in the difference spectrum 
		~that is in excellent agreement with DFT  and (ii) presence of a pseudogap at \e\, that is portrayed by fitting the near \e\, region with a power law function. 
		The present work emphasizes the electronic origin and the role of the atomic modulation  in hosting the CDW state in the martensite phase of stoichiometric \nmg.
	\end{abstract}
	
	\maketitle
	
	In recent years, considerable research has focused on understanding the intriguing physical phenomena connected to the charge density wave (CDW) state~\cite{Wang2022,
		PanSDSharma2022,
		Jiang2021,
		Shi2021,
		Zong2019}. 
	~ CDW is a collective excitation with periodic lattice distortion or modulation  that often results in a pseudogap at \e~\cite{3McKenzie95,Gruner88,3Lee73} and has been observed in various chalcogenide systems~\cite{Walmseley20,3Dai14,3Kim06}. 
	~\nmg\,  is an intriguing Heusler alloy having topologically protected nontrivial spin structures, such as skyrmions~\cite{3Phatak16}, where the  nature of atomic rearrangements related to existence of a bulk CDW state in the martensite phase  
	~has been a topic of intense study over the past three decades, but remains unresolved until date~\cite{3DSouza12,3Brown02,3Righi06,3Singh14,Martynov92,3Bungaro03,ZayakJPCM03,3Shapiro07,3Zheludev,Chulist10,MariagerPRB14,Singh15,Dsouza_ss12,Fukuda09,PonsJAP05,3Kaufmann10,GrunerSR18,Obata23}. In addition, \nmg\, is of practical importance due to its large magneto-caloric effect~\cite{3Zhou04,3Marcos03} and magnetic field induced strain (MFIS) of approximately 10\%~\cite{3Sozinov02,3Murray00}, the latter of which has been correlated with its large magnetocrystalline anisotropy~\cite{CasoliJPM20} and low twinning stress in the low temperature martensite phase.  
	
	It was observed quite early on that the martensite phase is not a simple tetragonal distortion of the high temperature cubic (austenite) phase, rather the structure has a periodic modulation. Using powder neutron diffraction  Brown \textit{ et al.} could account for all the reflections using a 7-fold supercell (7M) with a commensurate wave vector [$q_{_{\rm{CDW}}}$]  of $\frac{3}{7}c^*$ i.e. 0.4286$c^*$~\cite{3Brown02}. The modulation results from a periodic shuffling of (110)  planes of the cubic phase along the [1\={1}0] direction. Later on, structural studies using high resolution x-ray diffraction (XRD)~\cite{3Singh14,3Righi06}  showed that the martensite phase has a sinusoidal modulation in the positions of all the three elemental constituents indicating formation of CDW with $q_{_{\rm{CDW}}}$ of 0.425$c^*$. Significantly, from inelastic neutron diffraction study~\cite{3Zheludev}, the martensite phase was reported to be distorted by transverse modulation with incommensurate wave vector close to that reported from XRD~\cite{3Singh14,3Righi06}
	that was attributed to electron-phonon interactions and anharmonic effects. 
	A  phason excitation was observed in the martensite phase of \nmg~ from the neutron scattering experiment that was related to a CDW state~\cite{3Shapiro07}. Evidence of modulation in the martensite phase of \nmg\, was also observed in electron and x-ray diffraction studies from the appearance of the satellite spots~\cite{Dsouza_ss12,Fukuda09,Martynov92} - as is well known for chalcogenide materials that exhibit CDW~\cite{3DiMasi95,Sarkar23}.   From a  ultraviolet photoemission spectroscopy (UPS) study~\cite{3DSouza12}, existence of CDW in the pre-martensite phase~\cite{Singh13} was shown on the surface of \nmg(100) that continued to exist also in the martensite phase. However,  UPS being a highly surface sensitive technique with inelastic mean free path (IMFP) of 5\,\AA, the presence of CDW could not be inferred for the bulk martensite phase and the role of modulation was not probed.
	~A  time-resolved  experiment identified a coherent phonon that was related to the amplitude of the modulated structure~\cite{MariagerPRB14}. From an \textit{ab initio} theoretical study, Bungaro \textit{et al.} revealed that the dynamical instability in the TA$ 2$ phonon mode is connected with the nesting of the Fermi surface~\cite{3Bungaro03}, which is regarded as the distinguishing feature of the CDW state. Another first-principles study by Zayak \textit{et al.} reported that the martensite phase is stabilized by modulation that showed a tendency to exhibit a pseudogap~\cite{ZayakJPCM03}; the modulated structure was close to that reported in Ref.~\onlinecite{Martynov92}. 
	
Structural studies show that anti-site defects and disorder are not present in  \nmg ~\cite{3Brown02,3Righi06,3Singh14}. The  different  nuclear scattering amplitudes of Ni, Mn, and Ga in neutron diffraction gave occupancies of 1 for all three atoms at their respective sites~\cite{3Brown02}. However, non-stoichiometry that can be induced in Ni-Mn-Ga by preparing specimens deviated from the 2:1:1 atomic ratio of stoichiometric \nmg, influences  its magnetic and structural properties as well as the transition temperatures~\cite{BuchelMetals21,Vasilev99,3Banik07,Banik09,Singh11,Banik08,Singh15a}. For example, while the martensite start temperature ($T_M$) is around 206~K for stoichiometric \nmg, it increases for Ni excess compositions to as large as 537~K for $x$= 0.35 in Ni$_{2+x}$Mn$_{1-x}$Ga~\cite{Vasilev99,3Banik07}. In contrast, the Curie temperature  (376 K for  $x$= 0) decreases with increasing Ni content and for $x$$\sim$0.2 becomes equal to $T_M$~\cite{3Banik07,Vasilev99}. 

	Turning to the structural properties, in contrast to stoichiometric \nmg\, where modulated structure is observed~\cite{3Brown02,3Righi06,3Singh14}, for non-stoichiometric   Ni$_{2.19}$Mn$_{0.88}$Ga$_{0.93}$ i.e. $x$= 0.19,  the structure is tetragonal and modulation is not observed~\cite{3Kaufmann10,3Banik07}. For this composition, a 14M nanotwin model of the adaptive martensite phase was suggested by Kaufmann \textit{et al.}~\cite{3Kaufmann10}.  Similarly, signature of nanotwins  from transmission electron microscopy (TEM) was obtained for non-stoichiometric compositions that exhibit martensite phase at room temperature~\cite{PonsJAP05}, whereas \nmg\, exhibits the martensite phase below  $T_M$  (=206~K)~\cite{DSouzamsf}. Our survey of literature indicates that in compositions where the martensite phase has non-modulated tetragonally distorted Bain transition, the adaptive phase model~\cite{3Kaufmann10} is applicable.  However, a relatively recent density functional theory (DFT) calculation shows that even for stoichiometric  \nmg\, the modulation originates from the nanotwin ordering~\cite{GrunerSR18}. The authors establish that the phonon softening in the cubic austenite phase initializes the movement of the lattice planes, which seamlessly results in a nanotwinned adaptive martensite phase. However, this proposition is not supported by the conclusions of a large number of theoretical and experimental studies discussed above~\cite{3Brown02,3Righi06,3Singh14,Martynov92,3Bungaro03,ZayakJPCM03,3Shapiro07,3Zheludev,Chulist10,MariagerPRB14,Singh15,3DSouza12,Dsouza_ss12,Fukuda09}. In particular, 
	~the nanotwin structure~\cite{3Kaufmann10}  was ruled out for stoichiometric \nmg\, from high resolution XRD study  based on inhomogeneous displacements of the different atomic sites and the presence of phason broadening~\cite{3Singh14}. In spite of the above mentioned studies indicating formation of CDW in  bulk \nmg~\cite{3Brown02,3Righi06,3Singh14,Martynov92,3Bungaro03,ZayakJPCM03,3Shapiro07,3Zheludev,MariagerPRB14,Chulist10,Singh15,3DSouza12,Dsouza_ss12,Fukuda09}, a very recent theoretical study on \nmg\, using quasiparticle self-consistent GW (QSGW) method supports the formation of the 14M nanotwinned phase and thus raises doubt about the existence of the periodic modulation and the  CDW state in this system~\cite{Obata23}.
	
	In this paper, using a combination of DFT and hard x-ray photoelectron spectroscopy (HAXPES), we investigate the bulk electronic structure of \nmg\, in order to 
	~settle the aforementioned disagreement in the literature. 
	~We perform DFT calculations utilizing the actual experimental structures~\cite{3Brown02,3Righi06,3Singh14}, unlike previous theoretical investigations that used only non-modulated  and model structures~\cite{3Fujii89,3Barman05, ZayakJPCM03,GrunerSR18,Obata23}.  We also conducted DFT calculations utilizing the model nanotwin structure~\cite{3Kaufmann10}  in order to compare with the modulated structures. Due to the development of high brilliance synchrotron sources working in the stable top-up mode~\cite{Shin21}, HAXPES has turned out to be a useful technique to probe the bulk electronic structure~\cite{Woicik16,Grayandothers,3Nayak12,Singh22, 3Sarkar21,3Sadhukhan19}.  Although  there are a few HAXPES studies on other Heusler alloys~\cite{Ye10,Ueda22,Sadhukhan23}, the only  work on Ni-Mn-Ga system is a comparison of two non-stoichiometric compositions~\cite{3Kimura13}.
	~Thus, this work  is not related to our current investigation on stoichiometric \nmg\, in the austenite (300~K) and martensite phase (50~K) employing HAXPES and also DFT. In fact, the HAXPES study of the valence band (VB), in particular near the Fermi level (\e), would not only throw light on the CDW state, but also act as the ``gold standard"  for the DFT results to ascertain which structure best characterizes the martensite phase of \nmg. 
	
	\begin{figure}
		\includegraphics[width=0.85\linewidth,keepaspectratio,trim={50 0 0 0 },clip]{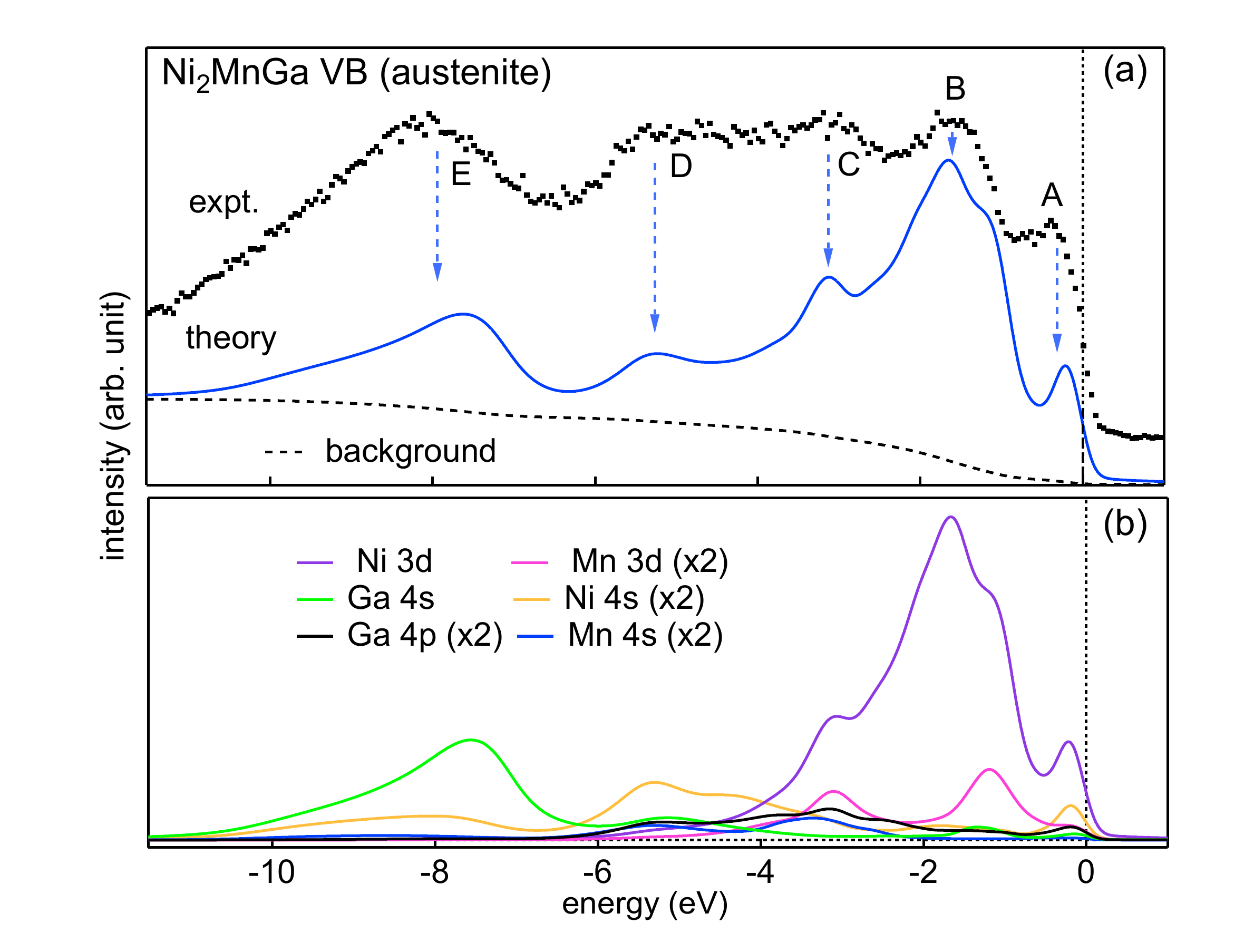}
		\caption{(a) The experimental HAXPES VB spectra of Ni$_2$MnGa  in the austenite phase taken with 6 keV photon energy at 300 K compared with the theoretical VB spectrum calculated from the partial DOS. Features \textbf{A} to \textbf{E} are marked by blue dashed arrows. The zero in the energy scale corresponds to the Fermi level (\e). 
			~(b) The partial atom and orbital projected components of  the theoretical VB.}
		\label{austenite}
	\end{figure}
	\vskip5mm
	\noindent\underline{\textit{HAXPES  and DFT VB spectra in the austenite phase:}}
	In Fig.~\ref{austenite}(a), we compare the theoretical and experimental VB for the austenite phase of \nmg\,  with L2$_1$ structure~\cite{3Webster84} [see Table~S1 for its structural parameters, also Fig.~S1 of the Supplementary material (SM)~\cite{supple} for the structure]. The HAXPES VB spectrum exhibits five distinct features  at about  -0.25 (\textbf{A}),  -1.6 (\textbf{B}),  -3.2 (\textbf{C}), -5.35 (\textbf{D}), and  -8~eV (\textbf{E}).  These features are the signatures of the  bulk electronic structure since  HAXPES is a bulk sensitive technique with an IMFP  of 66\,\AA\, (84\,\AA)  at 6 (8) keV~\cite{3tpp2m}. 
	
	In order to compare with the experiment, we have calculated the VB spectrum considering the $s$, $p$, and $d$ orbital projected components  of the  partial density of states (PDOS) of Ni, Mn, and Ga  considering their respective photoemission cross-sections~\cite{3Trz18}  (see the Methods section in SM~\cite{supple}).  In Fig.~\ref{austenite}(b), some of the dominant partial contributions to the calculated total VB spectrum [blue curve in Fig.~\ref{austenite}(a)]  are shown.  As shown by the blue dashed arrows in Fig.~\ref{austenite}(a), the energy positions of the features in the experimental spectra are in good agreement with the calculated VB spectra. 
	
	The sharp peak shown by \textbf{A} corresponds to the feature $a_1$, observed in the theoretical density of states (DOS), see Note~A and Fig.~S2 in SM~\cite{supple}. This arises due to the Ni 3$d$ minority states, with some admixture of the Ni 4$s$ states, with minor contribution from Mn and Ga states. The intense peak \textbf{B} is dominated by Ni 3$d$ majority and minority spin states, which corresponds to feature $c_1$ at -1.7 eV. A hump is observed at the lower binding energy side of \textbf{B}  [feature $b_1$] at about -1.1 eV. This has major contributions from Mn 3$d$ majority spin states and also from the Ni 3$d$ states in both the spin channels. Feature \textbf{C} is primarily due to the Ni 3$d$ states, with significant contributions from the Mn 3$d$ up states, additionally, Ga 4$p$ and Ni as well as Mn 4$s$ states also contribute. While \textbf{E} has a dominant contribution from Ga 4$s$, with some admixture of Ni 4$s$ states, \textbf{D} mainly arises from Ni 4$s$ states hybridized with Ga 4$s$, Mn 4$s$, and Ga 4$p$ states.  Additionally, a  peak  observed in the unoccupied  states around 1.45 eV (Fig.~S2 of SM~\cite{supple}) arises primarily from the minority spin Mn $3d$ states, whose  position is in good agreement with the inverse photoemission spectra~\cite{Maniraj_15}.
	
	A comparison of the HAXPES spectra taken with 8 and 6 keV shows that all the features \textbf{A-E} occur at similar energies (Fig.~S3 of SM~\cite{supple}). Note that larger photoemission cross-section of the $s$ states in HAXPES leads to appearance of features \textbf{C-E} in contrast to low energy phototemission such as x-ray photoelectron spectroscopy (XPS), where \textbf{C,D} are not visible and \textbf{E} is weak~\cite{DSouzamsf}.  In Fig.~S3 of SM~\cite{supple}, feature \textbf{B} appears at almost same energy in XPS and HAXPES. This indicates that the recoil effect~\cite{Fadley10} - a phenomenon observed in HAXPES of  light materials~\cite{Takata} as a shift of the photoemission peaks to higher binding energy - is not significant. In addition, as is the case for  \nmg,   the recoil effect has been reported to be insignificant for heavier 3$d$ transition metal systems~\cite{Sadhukhan23,3Nayak12,Singh22,3Sarkar21}.  
	
	It is noteworthy that according to a recent QSGW calculation~\cite{Obata23}, the austenite phase exhibits a peak right at \e\, in the minority spin DOS. This is in disagreement with our present (feature $a_1$ in Fig.~S2(a) of SM~\cite{supple}) 
	~as well as previous DFT results~\cite{3Fujii89,3Barman05,ZayakJPCM03,Ayuela99}. These DFT studies using the generalized gradient approximation (GGA) exchange-correlation functional (XC) observed that this peak appears between -0.19 to -0.22 eV which agrees nicely with feature \textbf{A} of the HAXPES VB. 
	~In light of the good agreement between results of DFT calculation performed with GGA XC and the  HAXPES data [Fig.~\ref{austenite}], it can be argued that the  GGA XC  quite accurately describes the electronic structure of  \nmg. This justifies its use for the in depth investigation of the martensite phase that has complicated  modulated structure with a large unit cell~\cite{3Brown02,3Righi06,3Singh14}. 
	
	\noindent\underline{\textit{VB spectra in the Martensite phase:}} In the literature, the first structural refinement of \nmg\ was carried out by Brown \textit{ et~al.}\cite{3Brown02} who reported  $q_{_{\rm{CDW}}}$ = $\frac{3}{7}c^*$ with  sinusoidal modulations for both Ni and Mn atoms, while Ga shows a non-sinusoidal modulation [see Table~S2 of SM~\cite{supple} and Fig.~8b of Ref.~\onlinecite{3Brown02}]. The structure is shown in Fig.~S1(b) and henceforth referred to as modulated-Brown (in short MDL-B). Righi \textit{et~al.} reported an incommensurate $q$ = 0.4248(2)$c^*$ [see Table~I of Ref. \onlinecite{3Righi06}] that can be approximated to a 7-fold supercell structure [see Fig.~S1(c) and Table~S3 of SM~\cite{supple} for structural parameters].  However, the authors estimated that their $q_{_{\rm{CDW}}}$ is closer to 0.4$c^*$ [= $\frac{2}{5}c^*$] and called it a 5M structure. In this paper, the name MDL-R has been attributed to this (modulated-Righi) structure. In contrast to MDL-B, in the MDL-R structure the amplitude and phase of modulation of all the atoms were similar (Fig.~S4  of SM~\cite{supple}). Moreover, it was  pointed out that the MDL-B structure contains some Ni-Mn and Ni-Ga distances -- 2.09 and 2.06\AA, respectively --  that are unexpectedly short~\cite{3Righi06}. Singh \textit{et~al.} found higher-order satellite reflections up to the third order and phason broadening of the satellite peaks in their XRD pattern and their refinement with the same super-space group as Righi \textit{et~al.} gave  $q_{_{\rm{CDW}}}$ = 0.4316(3)$c^*$, which was approximated to a 7-fold supercell structure with similar atomic modulations like the MDL-R structure and with a  $q_{_{\rm{CDW}}}$ value of $\frac{3}{7}$$c^*$ [see Table IV of Ref.~\onlinecite{3Singh14}]. This structure was referred to as 7M and is referred henceforth as modulated-Singh (MDL-S) structure (see Fig.~S1(d) and Table~S4 of SM~\cite{supple}).  Here, we have performed DFT calculation for all the three above discussed structures (MDL-B, MDL-R and MDL-S). We have also considered the nanotwin structure which comprises a periodic twinning, i.e. (5$\overline{2}$)$_2$, of the tetragonal non-modulated building blocks. Since this is a model structure, we have performed a full relaxation in our DFT calculation and this is referred to as nanotwin-Kaufmann (NTN-K) structure as shown in Fig.~S1(e) and Table~S5 of SM~\cite{supple}.  
	
	\begin{figure}[tb]
		\centering
		\includegraphics[width=0.9\linewidth,keepaspectratio,trim={50 0 0 0 },clip]{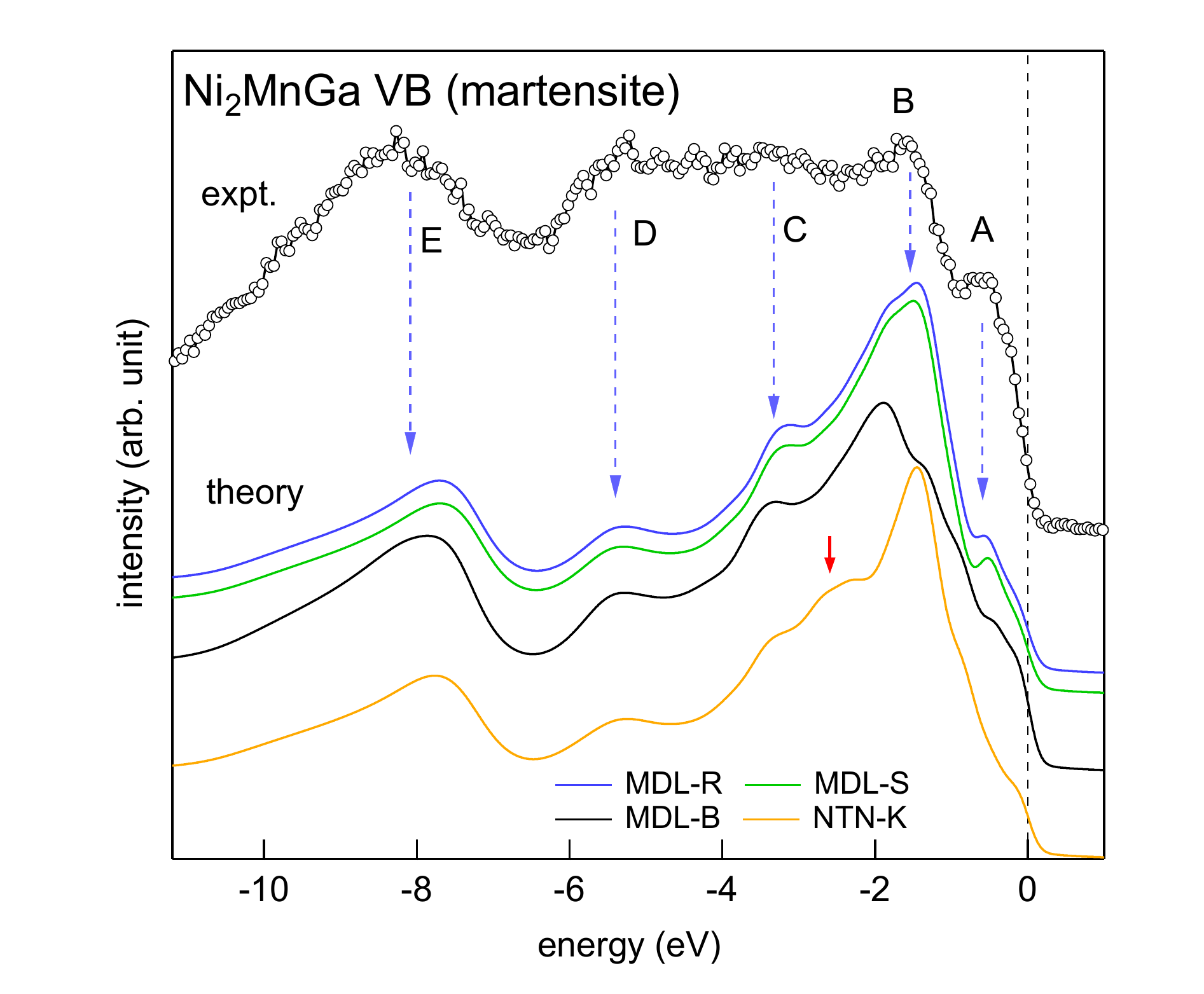}
		\caption{Theoretical VB spectra of Ni$_2$MnGa calculated for the different 7-fold modulated structures (MDL-R, MDL-S, and MDL-B) and the nanotwin 
			structure  (NTN-K). These are staggered along the vertical axis and compared with the experimental VB spectrum in the martensite phase at 50 K.}
		\label{martensite}
	\end{figure}
	
	\begin{table}[ht]
		\begin{center}
			\caption{Formation energy (in eV/atom) and magnetic moment values (in $\mu_B$ per f.u.) obtained from DFT calculations for different structures of \nmg\, in the martensite phase.}\vskip5mm
			\label{tab:1}
			\begin{tabular}{c c c c  c }
				\hline
				&  Structure  & Formation energy & Magnetic moment  
				\\
				\hline
				& MDL-B & -4.0325 & 3.94
				\\
				& MDL-R & -4.1783 & 4.21
				\\
				& MDL-S & -4.1927 & 4.20
				\\
				& NTN-K & -4.1837 & 4.20
				
				\\
				\hline
			\end{tabular}
		\end{center}
	\end{table}

	Table~1 shows that for all the structures although both the total energy and magnetic moment are rather close to each other, the MDL-S structure has the lowest energy showing that it is the most stable one. The  spin integrated total DOS (TDOS) and the PDOS for all the above mentioned structures -- as well as the calculated  VB along with some partial components -- are shown in Figs.~S5-S8  and discussed in Note B of SM~\cite{supple}. In Fig.~\ref{martensite}, we compare the calculated VB spectra for the above structures with the experiment.  The suppression and  shift of  feature \textbf{A} to -0.6 eV in the martensite phase compared to -0.25 eV in the austenite phase [Fig.~\ref{austenite}] in the HAXPES VB is nicely reproduced  by the MDL-R and MDL-S (MDL-R/S) structures [Fig.~\ref{martensite}], the feature \textbf{A} is over suppressed in MDL-B and is nearly absent in NTN-K. Further, we observe that the features  close to \e\, are dominated by the down spin states in case of the MDL-R/S structures, but for MDL-B and NTN-K cases, these have significant contributions from both the spin channels (Fig.~S2~\cite{supple}). Feature \textbf{B} is observed at the same energy position ($\sim$-1.5 eV as shown by the  blue dashed arrow) in all the structures except for MDL-B, where it is shifted considerably to about -2 eV. Thus the MDL-B structure does not show good agreement with experiment, and this could be related to unphysically short Ni-Mn and Ni-Ga distances~\cite{3Righi06}. While features \textbf{C-E} are well reproduced in all the modulated structures, NTN-K shows an extra feature  at -2.6 eV [red arrow], related to Ni 3$d$ minority spin states, that is absent in the experiment. Thus, overall the NTN-K structure does not show good agreement indicating that the  adaptive martensite model is not valid for stoichiometric \nmg. 
	
	From the above discussions, it is evident that the theoretical VB spectra based on the MDL-R/S structures are in very good agreement with the experimental results, and hence only these will be considered moving forward.  In fact, the VB spectra and the DOS of these two structures are quite similar  (compare Figs.~S6 and S7~\cite{supple}), which is related to the closeness of their crystal structure (compare Tables~S3 and S4 of SM~\cite{supple}). 
	
	We investigated the effect of electron-electron Coulomb interaction~\cite{Dudarev98} in \nmg\, using GGA+U calculations with the MDL-S structure considering  $U$ at both Mn  ($U_{\rm Mn}$) and Ni ($U_{\rm Ni}$) sites.  With a small $U_{\rm Mn}$ value of  0.5 eV and $U_{\rm Ni}$ = 0, we find that the agreement with feature \textbf{B} of the VB spectrum improves (Fig.~S9 of the SM~\cite{supple}). However, larger $U_{\rm Mn}$ and $U_{\rm Ni}$ values of 1.8 to $\sim$4 eV reported in literature~\cite{Janovec22,Koubsky18,Zeleny21} disagree with the experimental VB spectrum as well as the reported saturation magnetization values (4.04-4.27 $\mu_B$)~\cite{3Webster84,Devarajan13,Ooiwa92,Singh16} (see Note C and Figs.~S9 and S10 of the SM~\cite{supple}).  
		
		 In addition, although anti-site disorder is not reported in stoichiometric \nmg\, from diffraction studies~\cite{3Brown02,3Singh14,3Righi06}, we investigate the effect of 7\% anti-site disorder  by exchanging a Mn atom with a Ni or Ga atom in the MDL-S structure having 14 Mn atoms in its 56 atom unit cell. The 
			 DOS in Fig.~S11 of SM~\cite{supple} demonstrates that anti-site disorder does not alter the position of the features, although it does slightly broadens them, as was previously observed in other ternary materials~\cite{Sadhukhan23,Dsouza14}.

\begin{figure}[t]
		\includegraphics[width=0.9\linewidth,keepaspectratio,trim={0 0 0 0 },clip]{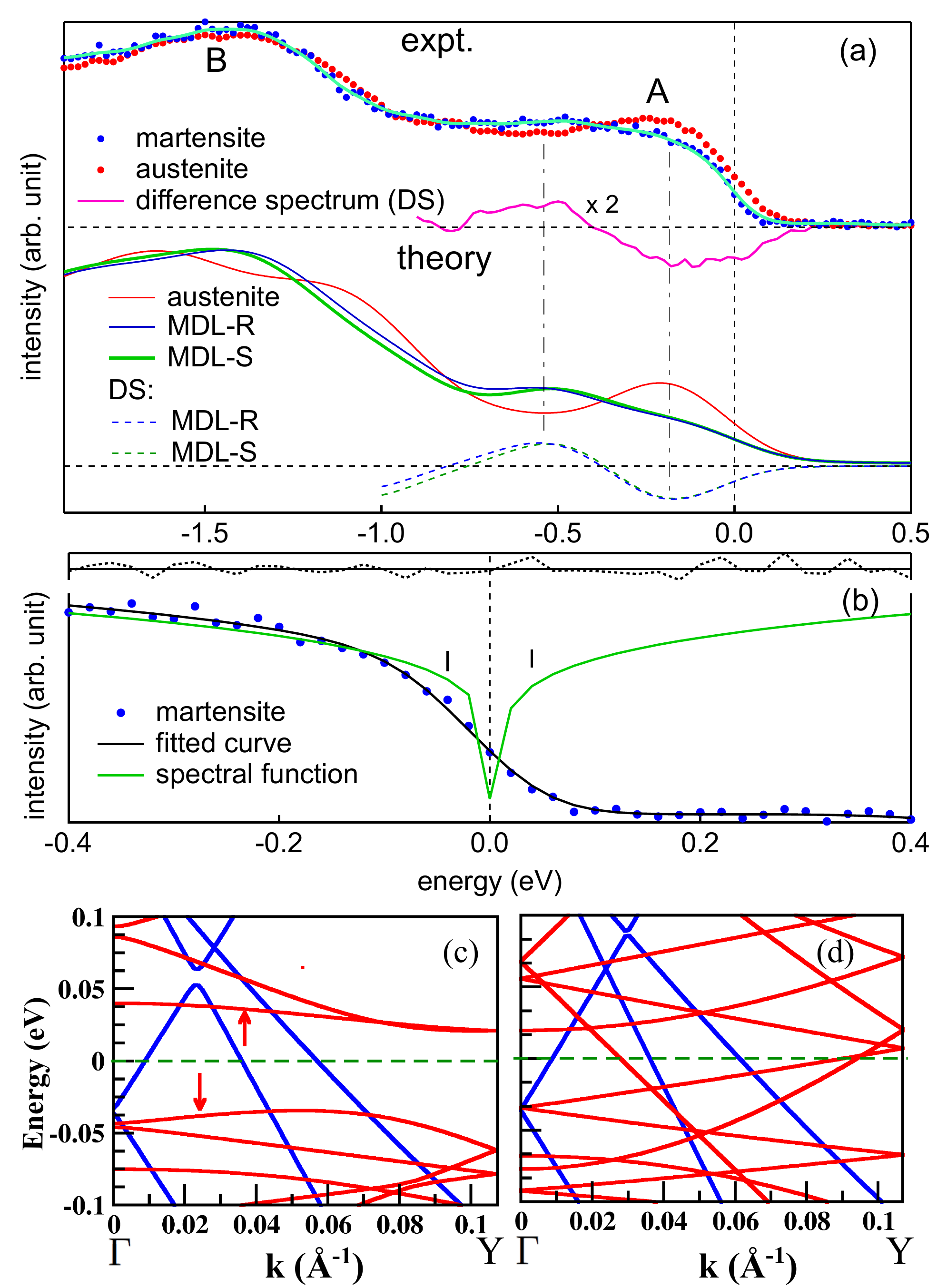}
		\caption{(a)  HAXPES VB of the martensite (50~K) and austenite (300~K) phases of \nmg~ in the near \e\, region (top panel) taken with small data steps.  The cyan curve represents the  martensite spectrum thermally broadened to 300~K. The bottom panel shows the theoretical VB spectra of the austenite, MDL-R/S  martensite structures.  Comparison of the dip and peak positions of the difference spectra (DS) between experiment (pink curve) and theory (dashed curves)  shown by the  dot-dashed vertical lines.  (b) The VB spectrum of the martensite phase around \e~(blue filled circles) fitted  (black curve) with a power law spectral function~\cite{function} (green). The residual is shown in the top panel. The  band structure (red: minority, blue: majority spin) along the \textbf{q}$_{_{\rm{CDW}}}$ for the  (c) MDL-S and  the (d)  non-modulated  NMDL-S structures.  }
\label{cdw}
\end{figure}

	\noindent\underline{\textit{CDW state in the martensite phase:}}
	The top panel of Fig.~\ref{cdw}(a) compares the experimental VB spectrum of the martensite and austenite phases. An interesting difference is observed between \e\, and  -1 eV:  in the martensite phase feature \textbf{A} is clearly suppressed,  while  an increased intensity around $\sim$-0.55~eV is observed in comparison to  the austenite phase. However, feature \textbf{B}, as well as features \textbf{C-E} from Figs.~\ref{austenite},\ref{martensite} do not exhibit any noticeable difference. This indicates that the states close to \e\, are  influenced by the phase transition. 
	
	As the thermal broadening of the Fermi function increases with temperature, we convoluted the low temperature (50~K) martensite spectrum with a Gaussian function of full width at half maximum of 4$k_B$$\Delta$$T$~\cite{3Chainani18} to obtain the difference spectra [DS = ($martensite-austenite$), top panel in Fig.~\ref{cdw}(a)] unaffected by this effect. Here $\Delta$$T$= $\sqrt{T_1^{2}-T_2^{2}}$,  $T_1$ = 300~K, $T_2$ = 50~K, and $k_B$ is the Boltzmann constant.  The  DS spectrum exhibits a dip centered around -0.1~eV and a peak at about -0.55~eV. This characteristic dip-peak shape points to a transfer of spectral weight from  -0.1~eV to -0.55~eV, which is known to be  a manifestation of the CDW state that has been observed in other  systems~\cite{3Dardel91,3Dardel92,3Matsuno01,3Yokoya05}.  In addition, the suppression of states at \e\, in the martensite phase is more than 30\%, indicating the creation of a pseudogap, which in turn also suggests a CDW state~\cite{3McKenzie95,3Lee73}. In order to confirm this, the shape of the spectrum near \e\, needs to be determined because, according to the theoretical formulation of CDW~\cite{3McKenzie95,3Balseiro80},  it should follow a power law function~\cite{function}  where $\alpha$ is the exponent. 
	This function was used to fit the near \e\, spectrum of the martensite phase using a least-square error minimization approach, where multiple starting values were applied and all parameters were adjusted, with the exception of the instrumental resolution. A random variation of the residual in the top panel of  Fig.~\ref{cdw}(b) shows that the fitting is satisfactory.

	Interestingly, the power law function portrays the pseudogap at \e, and its width is estimated to be 80 meV from the separation between the inflection points shown by the black ticks. $\alpha$ determines the shape of the spectral function, its value turns out to be 0.18$\pm$0.02,  which is close to that reported (0.16) for the surface CDW of \nmg\, in the martensite phase probed using surface-sensitive ($<$5\AA) UPS~\cite{3DSouza12}. This shows that the CDW has similar nature in the bulk and the surface. This is also supported by the similarity of $q_{_{\rm{CDW}}}$ 
	~value obtained from the surface sensitive low energy electron diffraction study~\cite{Dsouza_ss12} and the bulk value from XRD. 
	
	Having shown in Fig.~\ref{martensite} that the MDL-R/S  structures best describe the position of all the features \textbf{A-E}  of the experimental VB spectrum in the martensite phase, we examine whether the transfer of spectral weight is observed from DFT. In the lower panel of   Fig.~\ref{cdw}(a),  the theoretical DS for both  MDL-S and MDL-R show  excellent agreement -- as highlighted by the dot-dashed vertical lines -- in both position and shape of the dip and the peak compared to the experiment in the upper panel. 
	 Thus, the transfer of spectral weight is nicely depicted by the DOS from DFT. 
		
	In Figs.~\ref{cdw}(c,d), the band dispersion calculated along  \textbf{q}$_{_{\rm{CDW}}}$  i.e., $\Gamma$$Y$ 	for the  MDL-S structure  is compared to the non-modulated structure (NMDL-S). The amplitude of modulation  set to zero in NMDL-S, as shown in Fig.~S12 of SM~\cite{supple}.  Red arrows show an energy gap of 0.07-0.09 eV in  minority spin band of MDL-S. In contrast,  NMDL-S does not exhibit any gap.  This difference 
		~is thus directly related to the modulated CDW state.  The minority spin band that exhibits the gap  becomes relatively flat around -0.05 eV [Fig.~\ref{cdw}(c)]. This would result in an increase in the DOS around this energy and a decrease  closer to the \e\, that can explain the  transfer of spectral weight  [Fig.~\ref{cdw}(a)]. 	On the other hand,  the bands are nearly similar between MDL-S and NMDL-S along  other directions e.g.,  along $\Gamma$$X$ and $\Gamma$$Z$  and both spin bands cross \e\,   (Fig.~S13 of SM~\cite{supple}). This  results in a finite DOS at \e, (Fig. S2 of SM~\cite{supple}) indicating  presence of a Fermi edge in the photoemission spectrum. On the other hand, the pseudogap observed in the photoemission spectrum is attributed to a many-body effect, which includes the electron-phonon coupling~\cite{3Lee73} that is not taken into consideration in our calculation. 

	\noindent\underline{\textit{Conclusion:}} A combined experimental  and theoretical investigation of the bulk electronic structure of stoichiometric \nmg\, has been performed using HAXPES and DFT. The DFT calculations have been conducted for the modulated structures of the martensite phase as reported by previous diffraction studies, e,g.,  MDL-R/S~\cite{3Righi06,3Singh14} 
	~as well as the austenite phase with the L2$_{1}$ structure. Furthermore, the  nanotwin model structure (NTN-K)~\cite{3Kaufmann10} was considered. 
	A comparison of the theoretical VB spectra for the different martensite phase structures with HAXPES VB spectrum  reveals a very good feature to feature
	agreement in peak position and relative intensity for the MDL-R/S structures. 
	~In contrast, the NTN-K structure exhibits unsatisfactory agreement. This shows that the  modulation determined from diffraction studies~\cite{3Singh14,3Righi06} correctly describes the electronic structure of the martensite phase of \nmg.  A power law function fits the HAXPES VB close to \e\, revealing an 80 meV pseudogap.  Additionally, a transfer of spectral weight  occurs from the near \e\, region to the higher binding energy side, resulting in a dip-peak structure in the difference spectrum (DS).  A minority spin band exhibits a gap at \e\,  in the CDW state along the $q_{_{\rm{CDW}}}$ direction that can explain the transfer of spectral weight observed.  The pseudogap and the transfer of spectral weight  establish the existence of  CDW in the martensite phase of \nmg. The excellent agreement in the DS between experiment and the theory for the  MDL-R/S  structures show the role of the periodic atomic modulation in achieving the CDW state. Our calculations indicate that GGA is adequate for describing the electronic structure of \nmg\, if the correct structure is considered and large values of $U$ (1.8 to $\sim$4 eV) suggested recently~\cite{Zeleny21,Koubsky18,Janovec22}  contradict the experimental results. Our study establishes the electronic origin and the role of the atomic modulation  in hosting the CDW state in the martensite phase of stoichiometric \nmg\,  and resolves a recently generated controversy
	~\cite{GrunerSR18,Janovec22,Obata23}.
	
	\vskip5mm
	\noindent\underline{\textit{Acknowledgments:}} JB and AC thank the director, RRCAT for facilities and encouragement, and A. Banerjee and T. Ganguli for discussion. RRCAT computer division is thanked for the installation of the softwares and support. JB thanks RRCAT and HBNI for financial support. The HAXPES experiments were carried out at PETRA III of Deutsches Elektronen-Synchrotron, a member of Helmholtz-Gemeinschaft Deutscher Forschungszentren. Financial support by the Department of Science and Technology, Government of India within the framework of India@DESY collaboration is gratefully acknowledged. T. A. Lograsso and  D. L. Schlagel are thanked for providing us with the single crystal specimen. We are thankful to C. Schlueter and K. Biswas for support and encouragement. We would like to acknowledge the skillful technical support from  K. Ederer. 
	


\begin{thebibliography}{99}
\bibitem{function}
			$[w\times f(E,T) \times[|E -E_0|^\alpha]h(E - E_0) +\\
			|E^{\prime}_0 -E|^{\alpha}h(E^{\prime}_0 -E)]]\otimes (E,\sigma)$, \\
		where $E_0$ ($E^\prime$$_0$) is the threshold energy of the left (right) branch of the power law spectral function, $h$ denotes a unit step function, $\alpha$ is the exponent of the power law function, and $w$ is a multiplicative factor. $f(E,T)$ represents the Fermi function at temperature $T$.
\end{thebibliography}
\end{document}